\documentclass[12pt, onecolumn]{article} 

\usepackage{amsmath}  
\usepackage{amssymb}  
\usepackage{epsfig}  
\usepackage{wrapfig}  
\usepackage{array}  
\usepackage{axodraw}  

\makeatletter  
\renewcommand{\fnum@table}{\textbf{\tablename~\thetable}}  
\renewcommand{\fnum@figure}{\textbf{\figurename~\thefigure}}  
\makeatother  
  
\newlength{\myem}  
\newcounter{mysubequation}[equation]

\makeatletter  
\renewcommand{\section}{\@startsection{section}{1}{0em}%  
        {-3.5ex \@plus -1ex \@minus -.2ex}%   
        {2.3ex \@plus.2ex}%  
        {\normalfont\large\bfseries}}  
\renewcommand{\subsection}{\@startsection{subsection}{2}{0em}%  
        {-3.25ex\@plus -1ex \@minus -.2ex}%  
        {1.5ex \@plus .2ex}%  
        {\normalfont\bfseries}}  
\renewcommand{\subsubsection}%  
        {\@startsection{subsubsection}{3}{0em}%  
        {-3.25ex\@plus -1ex \@minus -.2ex}%  
        {1.5ex \@plus .2ex}%  
        {\normalfont\bfseries}}  
\makeatother

\begin{document}  
  
\thispagestyle{empty}  
\vspace*{.5cm}  
\noindent  
\hspace*{\fill}{\large OUTP-04/09}\\  
\vspace*{2.0cm}  
  
\begin{center}  
{\Large\bf TeV scale  resonant leptogenesis  
from supersymmetry breaking}  
\\[2.5cm]  
{\large Thomas Hambye, John March-Russell and Stephen M. West  
}\\[.5cm]  
{\it Theoretical Physics, Department of Physics\\  
University of Oxford, 1 Keble Road, Oxford OX1 3NP, UK}  
\\[.2cm]  
(March, 2004)  
\\[1.1cm]

{\bf Abstract}\end{center}  
\noindent  
We propose a model of TeV-scale resonant leptogenesis based upon recent 
models of the generation of light neutrino masses from supersymmetry-breaking
effects with TeV-scale right-handed (rhd) neutrinos, $N_i$.
The model leads to large cosmological lepton asymmetries via the resonant
behaviour of the one-loop self-energy contribution to $N_i$ decay.
Our model addresses the primary problems of previous phenomenological  
studies of low-energy leptogenesis: a rational for TeV-scale rhd
neutrinos with small Yukawa couplings so that the out-of equilibrium
condition for $N_i$ decay is satisfied; the origin of the tiny, but
non-zero mass splitting required between at least two  
$N_i$ masses; and the necessary non-trivial breaking of flavour  
symmetries in the rhd neutrino sector.  The low mass-scale of the   
rhd neutrinos and their superpartners, and the TeV-scale $A$-terms   
automatically contained within the model offer opportunities for  
partial direct experimental tests of this leptogenesis  
mechanism at future colliders.

\newpage  
  
\setcounter{page}{1}  
  
\section{Introduction}

The seesaw mechanism \cite{seesaw} and the associated mechanism of
leptogenesis \cite{FY} are  
very attractive means to explain the origin of the small neutrino   
masses and the baryon asymmetry of the universe. However they are  
very hard to directly test, and at present we have at best only weak  
circumstantial evidence that they are correct.  The primary difficulty  
is that the new particles that they involve have masses far beyond  
experimental reach.  In particular, in the standard realisation of the  
seesaw, the current lower bound for successful leptogenesis   
on the mass of the lightest right-handed neutrino $N_1$ is   
$M_{N_1} > 5 \cdot 10^{8}$~GeV   
\cite{DI,HMY,BDP,TH,GNRRS,HLNPS}.\footnote{This bound holds for  
  hierarchical rhd neutrinos assumed to be in thermal  
  equilibrium before decaying. It can be relaxed to   
  $M_{N_1} > 2 \cdot 10^{7}$~GeV if the lightest rhd neutrino is  
  assumed to be the dominant species populating the universe at the
  end of inflation  
  from the decay of the inflaton \cite{GNRRS}. It can also   
  be relaxed if the spectrum   
  of rhd neutrinos is not highly hierarchical \cite{HLNPS},   
  or if they are quasi-degenerate in mass \cite{Flanz,CRV,Pil1,TH,Pil2,HLNPS}.}  
Moreover the seesaw models contain many more parameters   
than there are low energy observables which could constrain them.   
For the standard seesaw-extended standard model with three right-handed  
neutrinos there are 18 parameters to be compared with 7 observables in  
the light neutrino mass matrix. (However,   
in its supersymmetric version with the non-trivial  
assumption of soft terms universality,  
rare lepton flavor changing and/or CP violating   
processes can give access to more parameters \cite{DI2,ER}.)  
  
Given this lack of direct evidence in favour of the standard  
mechanisms, it is important to consider possible low-energy, TeV-scale  
alternatives to, or variations of the standard seesaw leptogenesis  
mechanism, especially if they are testable. Additionally, such a  
mechanism could have the advantage of not requiring assumptions about  
the thermal history of the universe up to energy scales as high as  
$10^{10}$~GeV, as is usually the case, far above the temperature   
of the last epoch that has been tested, the epoch of nucleosynthesis  
at the MeV scale.  Moreover this would allow one to avoid the  
potential problems of the creation of dangerous relics,   
such as the gravitino, in too large a number at higher temperature.  
  
To build such a low-energy leptogenesis model is however a   
difficult task, essentially for the following reasons (for more details   
see \cite{TH}). First, low scale seesaw neutrino  
masses require tiny couplings and therefore generically induce too   
small a CP asymmetry.  Second, low scale means small Hubble constant which   
also requires tiny couplings in order that the decay of the particle at   
the origin of the asymmetry is not in thermal equilibrium.   
Third, a small   
Hubble constant requires in addition that the various scatterings which can  
suppress the asymmetry be under control. In particular, at such a low  
scale, the very fast gauge scatterings strongly prefer that
the decaying particle be a   
singlet of all low energy gauge symmetries.  
To avoid these problems, and limiting ourselves to standard thermal  
leptogenesis, one can think about three possibilities with decaying singlet  
particles: a large degeneracy of masses  
between the decaying particles \cite{Flanz,CRV,Pil1,TH,Pil2,HLNPS}; a   
hierarchy between the couplings of real and virtual heavy particles in  
the one loop leptogenesis diagrams (see \cite{TH}); or three body decays  
of the heavy particles with suppressed two body decays \cite{TH}.  
  
In this letter we will consider the first possibility that two or more  
particles are quasi degenerate.\footnote{Also in the context
of broken supersymmetric theories, Ref.\cite{bhs} considers an alternative
TeV-scale leptogenesis mechanism based upon hierarchical soft $A$-terms.}
In this case the asymmetry can be significantly enhanced  
through a resonant behaviour of the propagator of the virtual heavy  
particle in the leptogenesis self-energy diagram, Fig.~1.  As far as  
we are aware this is the only mechanism which can work at low  
scale within the standard leptogenesis model.    
However this framework still suffers from various significant  
difficulties: 1) As already mentioned above, at a scale as low  
as 1-10 TeV, neutrino  
mass constraints and the out-of equilibrium condition on the decay  
width require tiny Yukawa couplings of order $\sim 10^{-6}-10^{-7}$,  
and such small couplings need explanation.  
2) In order to compensate the large suppression of the asymmetry  
induced by these tiny couplings, an extremely tiny mass splitting  
is required  between two right-handed neutrino masses. The   
degree of degeneracy required has to be of order the value of the  
decay width to mass ratio, which implies that 
$(M_{N_1}-M_{N_2})/(M_{N_1}+M_{N_2}) < 10^{-10}$ \cite{Pil2}.
Such a tiny splitting, if not physically  
motivated, can be considered as a fine-tuning. 3) In a generic  
seesaw model there is no explanation why the right-handed neutrinos  
would have such a small mass ($M_N\sim$~TeV), just around the present  
experimentally reachable mass scale. 4) Finally, the tiny  
Yukawa couplings imply that the right-handed neutrino production  
cross sections are very suppressed, the right-handed neutrinos  
are not observable, and the model is not testable even if  
is located at a low scale.  Here  
we construct a resonant leptogenesis model  
which possesses a natural explanation of the first three  
problems and is in addition partially testable at future collider  
facilities such as the LHC.  It arises in the framework of the  
MSSM with L-violating soft supersymmetry breaking.  

Finally we note that recently a phenomenological Froggat-Nielsen   
type model has been proposed   
in \cite{Pil2} which gives splittings naturally of order the decay  
width and therefore leads naturally to a large asymmetry too.  Our   
model differs considerably from the one studied in Ref.\cite{Pil2}. 

%%%%%%%%%%%%%%%%%%%%%%%%%%%%%%%%%%%%%%%%%%%%%  
%  
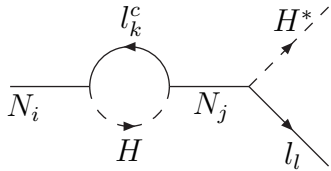
\begin{figure}[t]  
\begin{center}  
\begin{picture}(120,60)(0,0)  
%3rd diagram  
\Line(0,30)(30,30)  
\DashArrowArc(45,30)(15,180,360){5}  
\ArrowArc(45,30)(15,0,180)  
\Line(60,30)(90,30)  
\DashArrowLine(90,30)(120,60){5}  
\ArrowLine(90,30)(120,0)  
\Text(05,22)[]{$N_i$}  
\Text(45,6)[]{$H$}  
\Text(46,55)[]{$ l^c_k$}  
\Text(76,22)[]{$N_j$}  
\Text(107,55)[]{$H^\ast$}  
\Text(107,4)[]{$ l_{l}$}  
\end{picture}  
\end{center}  
\caption{Self-energy diagram for the right-handed neutrino decay.}  
\label{fig1}  
\end{figure}  
%%%%%%%%%%%%%%%%%%%%%%%%%%%%%%%%%%%%%%%%%%%%%  
  
\section{The model}

The most attractive suggested solution to the $\mu$ problem of  
supersymmetric theories is to invoke a symmetry to forbid the  
supersymmetry-preserving $\mu H_u H_d$ superpotential term; the  
presence of the $\mu$ term in the low-energy MSSM effective Lagrangian 
is explained as the result of an intermediate scale  
($m_I$) supersymmetry breaking expectation value within a   
higher-dimensional, $1/M_{\hbox{\footnotesize{Planck}}}$-suppressed  
operator \cite{GM}.  It has been recently   
emphasized \cite{gmnu,MW} that such a  
symmetry can also suppress the masses and interactions of the  
right-handed neutrinos in the same way (for related work see \cite{related}).    
This mechanism leads to TeV scale right-handed  
neutrinos with tiny Yukawa couplings which, in addition naturally  
lead to phenomenologically successful light Majorana neutrino masses  
\cite{gmnu,MW}, and an explicit and simple realization of  
this idea has been recently constructed \cite{MW}.   
  
The mechanism of Ref.\cite{MW} is based on the existence of two  
heavy standard model singlet fields: $X_{ij}$ which carries a  
flavour structure ($i,j=1,2,3$), and, $Y$, which is flavor
blind.\footnote{For
simplicity in this paper we simply assume 
flavour-blind $Y$ couplings.  This can result from a flavour
symmetry.  We leave the discussion of the various symmetry patterns
to a future publication \cite{west}.} In this model the various
particles have the   
following $R$-charge pattern: $X$ and $Y$ have charges $4/3$,   
the $N_i$ have charge $2/3$,  
the right-handed charged lepton superfields $E_i$  
have charge 2 and $H_u, H_d$ and $L_i$ all have $R$-charge  
equal to 0, and the usual $R$-parity is assumed, with in addition  
$R_p(X)=R_p(Y)=+1$.   
This leads to the following allowed $N_i$-dependent   
interactions in the superpotential  
\begin{equation}  
\mathcal{L}_N^W = \int d^2\theta \left( g\frac{X_{ij}}{M_P} L_i N_j H_u  
+ g^\prime\frac{Y}{M_P}L_i N_i H_u + \ldots \right),  
\label{newW}  
\end{equation}  
while the set of Kahler terms involving the   
right-handed $N_i$ fields are  
\begin{equation}  
\mathcal{L}_N^K = \int d^4 \theta \left(  
h  \frac{Y^\dagger}{M_P} N_i N_i +  
h_B \frac{Y^\dagger YX_{ij}^{\dagger} }{M_P^3} N_i N_j + 
h_x\frac{X_{ik} X_{kj}^\dagger Y^\dagger }{M_P^3} N_i N_j+\ldots \right),  
\label{newK}  
\end{equation}  
with $M_P = 1/\sqrt{8 \pi G_N} \sim 2 \cdot 10^{18}$~GeV the reduced  
Planck mass.  We will take both types 
of dimensionless couplings ($g$'s and $h$'s) to be of order one.  
In addition, we assume that after supersymmetry breaking in the
hidden sector, the   
field $Y$($X$) acquires a $F$($A$) component vacuum expectation  
value but no $A$($F$) component:  
\begin{equation}  
\langle Y \rangle_F=F_Y=f_Y m_I^2\,,\,\,\,\,  
\langle X_{ij} \rangle_A=A_{X_{ij}}=a_{X_{ij}} m_I \,,\,\,\,\,  
\langle Y \rangle_F=\langle X_{ij} \rangle_A=0\,,  
\label{FD}  
\end{equation}  
This then results in the following interactions of the softly  
broken right-handed neutrino-extended-MSSM (see e.g.~\cite{GR,NR}):  
\begin{equation}  
\mathcal{L}= \int d^2\theta \biggl( \lambda_{ij} L_i N_j H_u +  
M_N N_i N_i + \Delta M_{Nij} N_i N_i  \biggr) +   
A \tilde{L}_i \tilde{n}_i h_u +  
B^2_{ij} \tilde{n}_i \tilde{n}_j + \ldots,  
\label{lagr}  
\end{equation}  
where $\tilde{n}_i$ are the rhd sneutrino fields, $\tilde{L}_i$  
is the lhd slepton doublet, $h_u$ is the up-type  
Higgs scalar doublet, and the omitted terms include the usual  
soft scalar mass terms. In eq.~(\ref{lagr}) we have   
from eqs.~(\ref{newW}-\ref{FD})  
\begin{equation} 
\begin{array}{rclcl}  
\lambda_{ij} &=& g a_{Xij} (m_{3/2}/M_P)^{1/2}  &\sim & 10^{-7}-10^{-8},  
\\  
M_{N} &=& h f_Y m_{3/2} &\sim& 1 \, \hbox{TeV}\,,  \\  
\Delta M_{Nij} &=& h_x  a_{Xik} a_{Xkj}^\dagger f_Y m_{3/2}^2/M_P  
&\sim & 10^{-3} \, \hbox{eV} \,, 
\label{relations} \\ 
A &=& g^\prime f_Y m_{3/2} &\sim & 1 \, \hbox{TeV}\,,\\  
B^2_{ij} &=& h_B f_Y^2 a^s_{Xij} (m_{3/2}^{5}/M_P)^{1/2} &\sim &  
10^{-3} \, \hbox{GeV}^2   
\end{array}  
\end{equation}  
where the relations hold assuming $g$, $h$, $a_X$ and $f_Y$ to be of  
order unity with $m_{3/2}\simeq 1$~TeV, that is to say taking  
$m_I=\sqrt{m_{3/2} M_{P}} \sim 10^{11}$~GeV as we expect in hidden-sector  
models.\footnote{Note that there is no automatic vacuum stability 
problem implied by the $A$-terms.  The metastability condition \cite{vac}
requires that $|A|^2$ is smaller than the sum of lhd and rhd
sneutrino and up-like Higgs soft mass-squareds, and this can
easily be accommodated in our model by a modest suppression of $A$. 
Successful resonant leptogenesis puts no lower bound on $A$, while
the required neutrino masses can be accommodated by, eg, Suppressing
$A$ and enhancing $B$ by factors of $O({\rm few})$ from the order-of-magnitude
values given in eq.~(\ref{relations}). The probability during the
lifetime of the Universe of a single tunneling event to the true
vacuum can be estimated to be P $\sim H^{-4} M_{3/2}^4 e^{-S_4}$,
where $S_4\sim 46 (M_{3/2}/A)^2$, see section 5.3 of
Ref.\cite{Alinde}. For reasonable values of the ratio $M_{3/2}/A$ this
probability is vanishingly tiny.}

In this Lagrangian since the Yukawa couplings $\lambda$ are of  
order $10^{-7}-10^{-8}$ for $M_N$ of order TeV,  
the seesaw induced neutrino masses will be in general of order $m_\nu  
\sim \lambda^2 v^2/M_N \sim 10^{-5}$-$10^{-3}$ eV. As argued in  
Ref.\cite{MW} this could  
eventually explain the solar data but is to small to explain the  
atmospheric data. However at the one loop level the $B$ term  
together with the large $A$ terms and gauge interactions induce  
neutrino masses naturally of order the atmospheric lower bound on  
neutrino masses for the heaviest neutrino $\nu_3$, $m_{\nu_3} >  
\sqrt{\delta m^2_{\hbox{\footnotesize{atm}}}} \sim 0.05$~eV.  
This one loop contribution to the neutrino masses is of size   
\begin{equation}  
m_\nu^{\rm loop} \sim \frac{\alpha_w}{ 96 \pi} \frac{m_I^9 v^2}{M^5  
m_{\rm susy}^5} \simeq 10^{-2} \hbox{eV} - 10^{-1} \hbox{eV} \, ,  
\label{magnitude}   
\end{equation}  
and has flavour structure set by the lepton-number violating $B$-term mass,
$B_{ij}^2$, for the sneutrinos as given in eq.~(\ref{relations}), see
Refs.\cite{gmnu,MW}.  
Since the resulting light neutrino mass matrix is of exactly  
the same symmetric Majorana structure as in usual supersymmetric  
see-saw models, the counting of the Dirac and Majorana phases   
in $m_\nu^{\rm loop}$ is identical.

Note that since the $Y$ field is flavour blind the spectrum  
of right-handed neutrinos is degenerate at leading order, with only 
tiny splittings of order $10^{-15}-10^{-17}$ generated
by the higher-order term $X_{ik} X_{kj}^\dagger Y^\dagger N_i N_j$. 
This term comes from two contributions at the same order in $1/M_P$:
the tree level $h_x$ contribution of eq.(\ref{newK}) and the one-loop
contribution of Fig.2 induced by the Yukawa couplings, $\lambda_{ij}$.
These contributions give\footnote{For simplicity   
we have not performed a RG resummation of the logarithms in  
eq.~(\ref{beta}). For an example of such a procedure in the context of 
resonant leptogenesis, see Ref.\cite{Kt}.}  
\begin{eqnarray}  
M_{Nij}^{\rm corrected} &=& M_N\bigl[\mathbb{I}_{ij} + 
\beta \bigl(a_X a_X^\dagger + a_X^* a_X^T\bigr)_{ij}\bigr] \\ 
{\rm with~~}\beta &\sim& \frac{m_{3/2}}{h M_P} \biggl(h_x + 
\frac{g^2}{16 \pi^2} \log{\frac{M_P}{M_N}} \biggr) \sim 10^{-15} . 
\label{beta} 
\end{eqnarray} 
It is also important to stress that if there exists any other hidden sector 
flavour non-singlet field, $Z_{ij}$, with symmetry properties 
different from $X_{ij}$, then irrespective of 
these symmetry properties, the additional Kahler term   
\begin{equation}  
\mathcal{L} \owns \int d^4\theta \frac{1}{M^3_{\hbox{\footnotesize{P}}}}  
h_z Z_{ik} Z_{kj}^\dagger  Y^\dagger N_i N_j \,+\,  
 \mathcal{O}(1/M^4_{\hbox{\footnotesize{Planck}}})  
\label{lagrZa}  
\end{equation}  
cannot be forbidden.  The addition of such a term does not in any way  
disturb the nice properties of the above model for neutrino masses.  
However, if $Z_{ij}$ (similarly to $X_{ij}$) has no F-term but a $A$ term,   
$\langle Z_{ij} \rangle_A=a_{Z_{ij}} m_I$, an additional low-energy  
effective interaction emerges:  
\begin{equation}  
\mathcal{L} \owns \frac{m_{3/2}^2}{M_{P}}  
h_z a_{Z_{ik}} a_{Z_{kj}}^\dagger  f_Y^\dagger \int d^2\theta  
N_i N_j \,+\,  \mathcal{O}(1/M^2_P).  
\label{lagrZ}  
\end{equation}  
This results in a total rhd neutrino mass matrix of the form 
\begin{equation}  
M_N^R=M_N\bigl[\mathbb{I} + \beta(a_X a_X^\dagger + a_X^* a_X^T)  
+ \gamma(a_Z a_Z^\dagger + a_Z^* a_Z^T)\bigr]  
\label{massmatrix}  
\end{equation}  
with $M_N$ and $\beta$ given by eqs.~(\ref{relations}) and (\ref{beta}), 
and  
\begin{equation}   
\gamma = \frac{h_z f_Y}{h} \frac{m_{3/2}}{M_{\hbox{\footnotesize{P}}}} 
\sim 10^{-15} .
\end{equation}  
The $\beta$- and $\gamma$-dependent terms are irrelevant for
neutrino masses, but as we will  
show in the next section can be relevant for leptogenesis.

\section{Resonant leptogenesis}  
  
From the Lagrangian of eqs.~(\ref{lagr}) and (\ref{lagrZ}) we   
can now analyze the mechanism of leptogenesis that results.  
First, since the $A$ terms are of order $m_{3/2}$, that is to say of  
order unity at the $M_{\tilde{N}_i}$ scale, they will put the  
right-handed sneutrinos in deep thermal equilibrium.  Therefore the  
decay of the sneutrinos cannot lead to the creation of a large  
asymmetry.  
The right-handed neutrinos on the other hand have tiny effective Yukawa  
couplings of order $10^{-7}-10^{-8}$ and therefore will be naturally out of   
equilibrium, independent of the situation for the sneutrinos.  
In addition, the small Yukawa couplings of the right-handed  
neutrinos implies that the usual leptogenesis vertex
diagram leads to a far too  
small asymmetry (i.e.~$\varepsilon_{N_i} \sim \lambda^2/8 \pi  
< 10^{-15}$), and therefore it can be neglected.  
On the other hand, the self-energy diagram of Fig.1 for the rhd  
neutrinos, although also suppressed by Yukawa couplings, can be  
enhanced by a resonance effect if the mass splittings are naturally  
tiny. The $N_i$ asymmetry in this case is \cite{Pil1,Pil2,HLNPS}:  
\begin{equation}  
\varepsilon_i=-\sum_{j \neq i}  
  \frac{M_i }{M_j }\frac{\Gamma_j }{M_j }  
  I_{ij} S_{ij}  \,, \label{epsN}  
\end{equation}  
where  
\begin{equation} 
I_{ij} = \frac{ \hbox{Im}\,[ (\lambda  \lambda ^\dagger)_{ij}^2 ]}  
{|\lambda \lambda ^\dagger |_{ii} |\lambda \lambda ^\dagger |_{jj}}  
 \, ,\qquad  
S_{ij} = \frac{M^2_j  \Delta M^2_{ij}}{(\Delta M^2_{ij})^2+M_i ^2  
   \Gamma_j ^2} \, ,\qquad  
\Gamma_j = \frac{|\lambda \lambda ^\dagger |_{jj}}{8\pi}M_{j}  
\,.  
\label{selfenergy}
\end{equation}  
In the model of eqs.~(\ref{lagr}) and (\ref{lagrZ}), the    
lowest order $1/M_{P}$ contribution to the right-handed neutrino  
masses is flavor blind, eq.~(\ref{relations}), so the right-handed  
neutrino splittings vanish, which in turn leads to a  
vanishing asymmetry.  
However, as we explained above, at the next order in
$1/M_{\hbox{\footnotesize Planck}}$, there  
are two sources of mass degeneracy breaking, eq.(\ref{massmatrix}),
one from the term 
$X_{ik} X_{kj}^\dagger Y^\dagger N_i N_j/M_P^3$, and a second 
contribution of the same, small size, but with in general different 
flavour structure, from the $Z_{ij}$-dependent 
term of eq.~(\ref{lagrZ}).

It is important that the tiny mass splittings thus induced among 
$M_{N_i}$ mass eigenstates are of the same parametric size as 
the Yukawa-coupling-induced decay width of these massive states.  
Thus the propagator of the virtual rhd neutrinos in the self  
energy diagrams, eq.(\ref{selfenergy}), will
be naturally at the resonance or close to it.  
In more detail both contributions lead generically to mass 
splittings and decay widths of order
\begin{equation}
\Delta M_{ij}^2 \sim  m_{3/2}^2 \frac{m_{3/2}}{M_P}  \qquad {\rm and} \qquad
\Gamma_i \sim  \frac{m_{3/2}}{8\pi} \frac{m_{3/2}}{M_P}
\label{orders}
\end{equation} 
which therefore leads to  
\begin{equation}
S_{ij} \sim \frac{M_P}{m_{3/2}} \sim \frac{M_i}{8\pi\Gamma_i}
\end{equation}
and therefore to a possible total asymmetry of size
$\varepsilon_{i}\sim I_{ij}/8\pi$
which can be as large as $1/8\pi$ assuming there
are non non-trivial cancellations within $I_{ij}$ (see \cite{HLNPS}).  
To address this last issue it is necessary to   
diagonalise the $M_N^R$ mass matrix and to calculate  
the corresponding asymmetry in the mass eigenstates basis.  
To this end let us consider the two cases,  
$\gamma=0$, and $\gamma \neq 0$.  

%%%%%%%%%%%%%%%%%%%%%%%%%%%%%%%%%%%%%%%%%%%%%  
%  
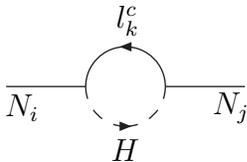
\begin{figure}[t]  
\begin{center}  
\begin{picture}(120,60)(0,0)  
%1st diagram  
\Line(0,30)(30,30)  
\DashArrowArc(45,30)(15,180,360){5}  
\ArrowArc(45,30)(15,0,180)  
\Line(60,30)(90,30)  
%\DashArrowLine(90,30)(120,60){5}  
%\ArrowLine(90,30)(120,0)  
\Text(06,22)[]{$N_i$}  
\Text(45,6)[]{$H$}  
\Text(46,55)[]{$ l^c_k$}  
\Text(85,22)[]{$N_j$}  
%\Text(107,55)[]{$H^\ast$}  
%\Text(107,4)[]{$ l_{l}$}  
\end{picture}  
\end{center}  
\caption{Self-energy contribution to eq.~(\ref{massmatrix}).}  
\label{fig2}  
\end{figure}  
%%%%%%%%%%%%%%%%%%%%%%%%%%%%%%%%%%%%%%%%%%%%%    

%%%%%%%%%%%%%%%%%%%%%%%%%%%%%%%%%%%%%%%%%%%%%%%%%%%%%%%%%%%%%%%%%%%%%%  
\section{Single source of flavour breaking: The $\gamma=0$ case}  
  
This pattern gives mass splittings naturally at the resonance as  
explained above but nevertheless turns out to result in  
$I_{ij}=0$ leading to a vanishing asymmetry.  This can be seen in  
the following way: Since the mass matrix of Eq.(\ref{massmatrix}) is   
real and symmetric it can be    
diagonalised by a real orthogonal matrix $O$ giving $O[a_X  
a_X^\dagger+(a_X a_X^\dagger)^T]O^T=D$ where $D$ is a real   
diagonal matrix. Furthermore writing $O(a_X a_X^\dagger)O^T \equiv   
D/2+C$ it is simple to check that $C$ is a purely imaginary  
antisymmetric matrix.  The one-loop mass eigenstate Yukawa 
couplings, $\lambda^{(1)}$, are therefore related to the tree 
level couplings, $\lambda$, via a flavour rotation 
$\lambda^{(1)}=O \lambda$.  Upon substitution of the 
mass-eigenstate basis Yukawa's into the $I_{ij}$ of eq.(\ref{epsN}) we 
obtain $I_{ij} \sim
\hbox{Im}[(\lambda^{(1)} \lambda^{(1) \dagger})_{ij}^2] \propto 
\hbox{Im}[(O a_X a_X^\dagger O^T)_{ij}^2]
\propto \hbox{Im}[(D/2+C)_{ij}^2]=0$, for $i\neq j$.  
Thus we learn that a single source of flavour breaking as contained in the  
Yukawa couplings can lead to non-trivial mass splittings among the $N_i$,  
but does {\em not} lead to any CP-violation.  
Intuitively this is not too surprising since the self energy diagram of  
Fig.~1 is the same as the one in the leptogenesis self-energy diagram,   
Fig.~2.  
  
Note that, in the context of the ordinary see-saw extended SM,
the possibility of one-loop induced resonant leptogenesis    
has been considered in Ref.~\cite{lisboa}, for a  
phenomenological case with two right-handed neutrinos with  
$M_{N_1}=M_{N_2}$ but different Yukawa couplings, and it
was claimed that this leads to a naturally large resonantly enhanced   
asymmetry. However in this calculation it appears that the numerator  
of the asymmetry has been calculated with the tree level Yukawa  
couplings. Since the full, one-loop corrected masses differ from  
the tree level ones only by a small amount, one might   
expect that it is a good approximation not to include the  
full one-loop correction that arises from going to the mass
eigenstate basis.  However since the tree level masses  
are exactly degenerate, even a small off-diagonal contribution  
leads to large mixing and therefore it is necessary to  
renormalise as we have done here.  We then see that in fact  
this system gives a vanishing asymmetry.  
  
We conclude from the above study of the $\gamma=0$ case that  
degenerate rhd neutrinos with different Yukawa couplings  
do not lead to a one-loop self-energy leptogenesis contribution. 
For the resonant leptogenesis  
mechanism to work it is essential to have at least two sources of  
flavor breaking, so we now turn to the analysis with $\gamma\neq 0$.

%%%%%%%%%%%%%%%%%%%%%%%%%%%%%%%%%%%%%%%%%%%%%%%%%%%%%%%%%%%%%%%%%%%%%  
\section{Two sources of flavour breaking: The $\gamma \neq 0$ case}  
  
When $\gamma \neq 0$ in eq.(\ref{massmatrix}) a second source of  
flavour structure is introduced, and  
as we will argue in this Section, it can give $I_{ij}$ factors which  
can be as large as $1$ and $S_{ij}$ factors naturally close to  
the resonance, that is to say of order  
$M_{N_i}/2\Gamma_{i}$, leading to CP asymmetries of order one.  
  
To convince oneself that the CP-asymmetry can now be non-zero, it
is useful to choose a special form for the three by three
$\langle Z\rangle = a_{Z_{ij}}m_I$ matrix in eq.~(\ref{massmatrix})
such that the off-diagonal terms in the total mass matrix of
eq.(\ref{massmatrix}) cancel between the $\beta$- and $\gamma$-dependent
terms, but leaving non-trivial diagonal entries $M_N^R={M_N}\mathbb{I} +  
\hbox{diag}(\delta M_{N_1}, \delta M_{N_2},\delta M_{N_3})$ where
$\delta M_{N_1}< \delta M_{N_2} < \delta M_{N_3}$. The $\delta{M_{N_i}}$  
are naturally of size $m_{3/2}(m_{3/2}/M_{\hbox{\footnotesize{P}}})$, and
can be chosen to be positive, leading to a mass splitting
of the same order.\footnote{This example
requires a tuning between both contributions of  
eq.~(\ref{massmatrix}) but it is convenient to demonstrate that   
a large asymmetry can be obtained.}  
Since the total mass matrix is a diagonal matrix the Yukawa
couplings in this  
case do not need to be rotated to the mass eigenstate basis
and $I_{ij}$ is therefore just a function of the lowest order
Yukawa coupling matrix $\lambda_{ij}$.  
In addition, one can re-express the total asymmetry as
\begin{equation}
\varepsilon_{\rm tot}=-A_{21} I_{21}-A_{31} I_{31}-A_{32} I_{32}
\label{eptot}
\end{equation}
where $\varepsilon_{\rm tot}=\sum_{i}\varepsilon_i$, 
the $I_{ij}$'s are defined in eq.~(\ref{selfenergy}), and
with all $A_{ij}$'s in eq.(\ref{eptot}) being the positive quantities:  
\begin{equation}
A_{ij}=\frac{M_iM_j}{8 \pi}\Delta M^2_{ij}\left[ \frac{|\lambda \lambda
  ^\dagger |_{jj} }{(\Delta M^2_{ij})^2+M_i ^2\Gamma_j
  ^2}+\frac{|\lambda \lambda ^\dagger |_{ii} }{(\Delta M^2_{ij})^2+M_j
  ^2\Gamma_i ^2} \right] \, .
\label{Aij}
\end{equation}
It is now easy to see that one can
choose a form for the Yukawa couplings such that, eg,
$I_{21}=I_{31}=0$ while $I_{32}\neq 0$ and $A_{32}\neq 0$.  Thus
$\varepsilon_{\rm tot}$ is non zero in this case. 
This completes the existence proof that $\gamma\neq 0$ can lead to 
non-zero asymmetries as claimed.

Moreover, general forms for the Yukawas and $\langle
Z_{ij}\rangle_A$'s will produce 
independent $A_{ij} I_{ij}$ terms and consequently give rise to a non
zero asymmetry, and applying the relations displayed
in eq.~(\ref{orders}) we get,
\begin{equation}
\varepsilon_{\rm total} \sim -\frac{1}{8\pi} (aI_{21}+bI_{31}+cI_{32}) .
\label{Sestim}
\end{equation}
where $a,b,c$ are $\cal{O}$(1) coefficients arising from the self-energy
terms in eq.(\ref{Aij}). 
A straightforward numerical investigation of the dependence of
eq.(\ref{Sestim})  on the forms and phases of the effective Yukawa
matrix $\lambda_{ij}$, together with
the $\alpha$ and $\beta$ terms of the mass
matrix eq.~(\ref{massmatrix}), shows that it is simple to find cases
where $aI_{21}+bI_{31}+cI_{32}\sim 1$\footnote{Note,
that unlike the usual see-saw model based leptogenesis case, one
cannot re-write the $I_{ij}$'s in terms of the 
solar and atmospheric neutrino mass-squared
differences, mixing angles, and phases in the conventional
way.  The reason for
this is that the light neutrino spectrum is dominantly set
by the one-loop contribution of eq.~(\ref{magnitude}) which depends upon the
$B$-term susy- and lepton-number-breaking mass of the sneutrinos, with only
small corrections (possibly leading to the small
$\Delta m^2_{\rm solar}/\Delta m^2_{\rm atm}$ light neutrino hierarchy)
arising from a tree-level see-saw contribution depending upon $\lambda$
in the usual way, see Ref.\cite{MW}.}.

Furthermore, for natural values of the model parameters, the produced  
lepton asymmetry is not suppressed by any wash-out effect.  
The Yukawa couplings which are of order $10^{-7}-10^{-8}$ give   
a decay width smaller than the Hubble constant and therefore will  
not induce any wash-out effect via decay or via  
scatterings, neither in the Boltzmann equation of the $N_i$ number  
density nor in the Boltzmann equation of the lepton number density.  
Moreover the $A$-terms, even if large, $A\sim$~TeV, cannot  
change this result because they can effect the $N_i$ number densities  
only via scatterings which are also suppressed by Yukawa couplings,  
and they do not affect the lepton number density because  
they can break lepton number only when accompanied by Yukawa couplings  
or by a $B$ term which is also very suppressed.  
In particular, it can be checked that the potentially dangerous  
$\tilde{L}+H \leftrightarrow \tilde{N} \leftrightarrow \tilde{L}^\ast +  
H^\ast$ process induced by two $A$ terms and a $B$ term has in fact a   
rate smaller than the Hubble constant.  
Thus the produced $n_L/s$ can be naturally of order  
$\varepsilon_{N_i}/g_\star\sim \varepsilon_{N_i}/100$  
and therefore from the above discussion can be naturally as   
large as $1/(100\cdot 8\pi)$ for order one couplings!   
To our knowledge, this is the only model of thermal leptogenesis which  
can lead naturally to such a large asymmetry.  
Note however that the asymmetry can be rapidly suppressed if we allow  
the phase to be non-maximal and the couplings to be not all of order  
unity.  Any deviation of the constants $h$, $a_X$, or $a_Z$ by
one order of magnitude   
in eq.(\ref{Aij}) can lead    
to several orders of magnitude suppression of the asymmetry.  
This can lead easily to the CMBR-determined experimental value:  
$n_B/n_\gamma=6.1^{+0.3}_{-0.2} \cdot 10^{-10}$ \cite{WMAP}.  
  
What about the testability of our model? The main  
attractive point here is that since we have large $A$  
terms the sneutrinos could be observed quite easily  
and allow one to test to a large extent the one-loop diagrams
at the origin of the neutrino mass.  
From the additional discovery of supersymmetry we could in further  
conclude that TeV-scale right-handed neutrinos also must exist.  
Although the later could not be observed because they can be produced  
only by the tiny Yukawa couplings, the primary ingredients of our TeV-scale
leptogenesis mechanism could thus be tested.

\section{Conclusions}  
  
In summary, in the framework of softly broken supersymmetric theories we   
have proposed a natural model of resonant leptogenesis utilizing  
TeV scale rhd neutrinos, $N_i$.  Our discussion  
is based upon recent models of the generation of light neutrino  
masses from supersymmetry-breaking effects, which provide a  
natural explanation for the presence of rhd neutrinos  
at the TeV scale, and for the small Yukawa couplings and  
$B$-terms necessary for the correct resulting light neutrino  
spectrum.  We show that these properties also lead to  
large cosmological lepton asymmetries, via the resonant behavior  
of the one-loop self-energy contribution to $N_i$ decay. The model  
addresses the primary problems of previous phenomenological  
studies of low-energy leptogenesis: a rational for small Yukawa couplings  
so that the out-of equilibrium condition on the $N_i$ decay is satisfied;  
the origin of tiny, but non-zero mass splitting required between at least two  
$N_i$ masses; and the necessary non-trivial breaking of flavour
symmetries in the rhd neutrino sector.  The low mass-scale of the  
rhd neutrinos and their superpartners, and the weak-scale $A$-terms 
automatically contained within the model, offer opportunities for 
partial direct experimental tests of this leptogenesis 
mechanism at future colliders.

%\vskip 0.05in 
\begin{center} 
{\bf Acknowledgments} 
\end{center} 
%\vskip0.05in 
We wish to thank Lotfi Boubekeur, Yin Lin, Michele Papucci 
and Goran Senjanovic
for discussions. T.H. is
supported by EU Marie Curie contract HPMF-CT-01765 and SW is 
supported by PPARC Studentship Award PPA/S/S/2002/03530.


\begin{thebibliography}{54} 
% 

\bibitem{seesaw} 
M. Gell-Mann, P. Ramond and R. Slansky, 
in {\it Supergravity}, edited by P. van Nieuwenhuizen and D. Freedman, 
(North-Holland, 1979), p.~315; 
S.L. Glashow, in Quarks and Leptons, Carg\`ese, eds. M. L\'evy et al., 
(Plenum, 1980, New-York), p. 707; 
T. Yanagida, in {\it Proceedings of the Workshop on the Unified Theory 
and the Baryon Number in the Universe}, edited by O. Sawada and
A. Sugamoto (KEK Report No.~79-18, Tsukuba, 1979), p.~95; 
R.N.~Mohapatra and G. Senjanovi\'{c}, Phys. Rev. Lett. {\bf 44}, 
(1980) 912. 
 
\bibitem{FY}
M.~Fukugita and T.~Yanagida,
%``Baryogenesis Without Grand Unification,''
Phys.\ Lett.\ B {\bf 174}, 45 (1986).
%%CITATION = PHLTA,B174,45;%%

\bibitem{DI}
S. Davidson and A. Ibarra, Phys. Lett. {\bf B535} (2002) 25. 
 
\bibitem{HMY}
K. Hamagushi, H. Murayama and T. Yanagida,  
Phys. Rev. {\bf D65} (2002) 043512. 
 
\bibitem{BDP}
W. Buchm\"uller, P. Di Bari and M.~Pl\"umacher, arXiv:hep-ph/0302092;  
Phys. Lett. {\bf B547} (2002) 128;  
Nucl. Phys. {\bf B643} (2002) 367; Nucl. Phys. {\bf B665} (2003) 445;
arXiv:hep-ph/0401240. 
 
\bibitem{TH}
T.~Hambye, Nucl. Phys. {\bf B633} (2002) 171. 
 
\bibitem{GNRRS} 
G.F. Giudice, et al, arXiv:hep-ph/0310123. 
 
\bibitem{HLNPS}
T.~Hambye, et al, arXiv:hep-ph/0312203. 
 
  
\bibitem{Flanz}
M. Flanz, E.A. Paschos and U. Sarkar, Phys. Lett. 
{\bf B345} (1995) 248;  
M.~Flanz, et al, Phys. Lett. {\bf B389} (1996) 693.  
 
\bibitem{CRV}
L. Covi, E. Roulet and F. Vissani, Phys. Lett. {\bf B384} (1996) 169. 
 
\bibitem{Pil1}
A. Pilaftsis, Phys. Rev. {\bf D56} (1997) 5431; Nucl.  
Phys. {\bf B504} (1997) 61. 
 
\bibitem{Pil2} 
A. Pilaftsis and T.E.J. Underwood,  arXiv:hep-ph/0309342. 
 
\bibitem{DI2}
S. Davidson and A. Ibarra, JHEP {\bf 0109} (2001) 013. 
 
\bibitem{ER}
J. Ellis and M. Raidal, Nucl.Phys. {\bf B643} (2002) 229. 
  
\bibitem{bhs}
L.~Boubekeur, T.~Hambye and G.~Senjanovic,
%``Low-scale leptogenesis and soft supersymmetry breaking,''
arXiv:hep-ph/0404038.
%%CITATION = HEP-PH 0404038;%%


\bibitem{GM}
G.F. Giudice and A. Masiero, Phys. lett. {\bf B206} (1988) 480. 
 
\bibitem{gmnu} 
N.~Arkani-Hamed, et al,
%``Small neutrino masses from supersymmetry breaking,'' 
Phys.\ Rev.\ D {\bf 64}, 115011 (2001) 
[arXiv:hep-ph/0006312]; and
%%CITATION = HEP-PH 0006312;%% 
%``Neutrino masses at v**(3/2),'' 
arXiv:hep-ph/0007001: 
%%CITATION = HEP-PH 0007001;%% 
F.~Borzumati and Y.~Nomura, 
%``Low-scale see-saw mechanisms for light neutrinos,'' 
Phys.\ Rev.\ D {\bf 64}, 053005 (2001) 
[arXiv:hep-ph/0007018]: 
%%CITATION = HEP-PH 0007018;%% 
F.~Borzumati, et al,
%``Variations on supersymmetry breaking and neutrino spectra,'' 
arXiv:hep-ph/0012118. 
%%CITATION = HEP-PH 0012118;%% 

\bibitem{MW}
%\cite{March-Russell:2004uf}
J.~March-Russell and S.~M.~West,
%``A simple model of neutrino masses from supersymmetry breaking,''
arXiv:hep-ph/0403067.
%%CITATION = HEP-PH 0403067;%%J. March-Russell and S. West,
 %%arXiv:hep-ph/0403067. 


\bibitem{related}
Y.~Grossman and H.~E.~Haber,
%``Sneutrino mixing phenomena,''
Phys.\ Rev.\ Lett.\  {\bf 78} (1997) 3438 [arXiv:hep-ph/9702421]:
%%CITATION = HEP-PH 9702421;%%
F.~Borzumati, K.~Hamaguchi and T.~Yanagida,
%``Supersymmetric seesaw model for the (1+3)-scheme of neutrino masses,''
Phys.\ Lett.\ B {\bf 497} (2001) 259
[arXiv:hep-ph/0011141]:
%%CITATION = HEP-PH 0011141;%%
R.~Kitano,
%``Small Dirac neutrino masses in supersymmetric grand unified theories,''
Phys.\ Lett.\ B {\bf 539} (2002) 102 [arXiv:hep-ph/0204164]:
%%CITATION = HEP-PH 0204164;%%
J.~A.~Casas, J.~R.~Espinosa and I.~Navarro,
%``New supersymmetric source of neutrino masses and mixings,''
Phys.\ Rev.\ Lett.\  {\bf 89} (2002) 161801
[arXiv:hep-ph/0206276]:
%%CITATION = HEP-PH 0206276;%%
R.~Arnowitt, B.~Dutta and B.~Hu,
%``Yukawa textures, neutrino masses and Horava-Witten M-theory,''
arXiv:hep-th/0309033:
%%CITATION = HEP-TH 0309033;%%
S.~Abel, A.~Dedes and K.~Tamvakis,
%``Naturally small Dirac neutrino masses in supergravity,''
arXiv:hep-ph/0402287.
%%CITATION = HEP-PH 0402287;%%

\bibitem{west}
%\cite{West:2004me}
S.~M.~West,
%``Naturally Degenerate Right Handed Neutrinos,''
arXiv:hep-ph/0408318.
%%CITATION = HEP-PH 0408318;%%S.M.West, OUTP-04/16.

\bibitem{GR}
G.~D'Ambrosio, G.F. Giudice and M. Raidal, Phys. lett. {\bf B575} (2003) 75. 
   
\bibitem{NR}
Y. Grossman, et al, Phys. Rev. Lett. {\bf 91} (2003) 251801. 

\bibitem{vac}
J.~M.~Frere, D.~R.~T.~Jones and S.~Raby,
%``Fermion Masses And Induction Of The Weak Scale By Supergravity,''
Nucl.\ Phys.\ B {\bf 222} (1983) 11; 
L.~Alvarez-Gaume, J.~Polchinski and M.~B.~Wise,
%``Minimal Low-Energy Supergravity,''
Nucl.\ Phys.\ B {\bf 221} (1983) 495;
J.~P.~Derendinger and C.~A.~Savoy,
%``Quantum Effects And SU(2) X U(1) Breaking In Supergravity Gauge Theories,''
Nucl.\ Phys.\ B {\bf 237} (1984) 307;
C.~Kounnas, A.~B.~Lahanas, D.~V.~Nanopoulos and M.~Quiros,
%``Low-Energy Behavior Of Realistic Locally Supersymmetric Grand Unified
%Theories,''
Nucl.\ Phys.\ B {\bf 236} (1984) 438;
M.~Claudson, L.~J.~Hall and I.~Hinchliffe,
%``Low-Energy Supergravity: False Vacua And Vacuous Predictions,''
Nucl.\ Phys.\ B {\bf 228} (1983) 501;
J.~A.~Casas, A.~Lleyda and C.~Munoz,
%``Strong constraints on the parameter space of the MSSM from charge and color
%breaking minima,''
Nucl.\ Phys.\ B {\bf 471} (1996) 3
[arXiv:hep-ph/9507294].

\bibitem{Alinde}
A.D. Linde, {\it Particle physics and Inflationary Cosmology} (Harwood
Academic Publishers, Chur, Switzerland 1990).

%\cite{Turzynski:2004xy}
\bibitem{Kt}
K.~Turzynski,
%``Degenerate minimal see-saw and leptogenesis,''
Phys.\ Lett.\ B {\bf 589} (2004) 135
[arXiv:hep-ph/0401219].
%%CITATION = HEP-PH 0401219;%%

\bibitem{lisboa}
R. Gonzalez Felipe, F.R. Joaquim and B.M. Nobre,  arXiv:hep-ph/0311029. 
 
\bibitem{WMAP}
D.N. Spergel et al., arXiv:astro-ph/0302209. 

% 
%%%%%%%%%%%%%%%%%%%%%%%%%%%%%%%%%%%%%%%%%%%%%%%%%%%%%%%%%%%%%%%%%%%%%% 
% 
\end{thebibliography}
\end{document}